\documentclass[preprint,pra,superscriptaddress]{revtex4-1}

\usepackage{amscd}
\usepackage{amsthm,amsfonts,dsfont}
\usepackage{enumerate}

\usepackage{mathrsfs}
\usepackage{amssymb}
\usepackage{graphicx}
\usepackage{dcolumn}
\usepackage{bm}
\usepackage{textcomp}
\usepackage{amsmath}
\usepackage[titletoc]{appendix}

\usepackage{epsfig}
\usepackage{epstopdf}
\usepackage{subfigure}
\usepackage{color}
\usepackage[10pt]{moresize}

\newcommand\ket[1]{\ensuremath{|#1\rangle}}

\newcommand\oprod[2]{\ensuremath{|#1\rangle\langle#2|}}

\newcounter{RomanNumber}

\begin{document}
\title{Experimental Side-Channel-Free Quantum Key Distribution}

\author{Chi Zhang}
\affiliation{Hefei National Laboratory for Physical Sciences at Microscale and Department of Modern Physics, University of Science and Technology of China, Hefei 230026, China}
\affiliation{Shanghai Branch, CAS Center for Excellence and Synergetic Innovation Center in Quantum Information and Quantum Physics, University of Science and Technology of China, Shanghai 201315, China}
\affiliation{Jinan Institute of Quantum Technology, Jinan, Shandong 250101, China}

\author{Xiao-Long Hu}
\affiliation{State Key Laboratory of Low Dimensional Quantum Physics, Department of Physics, Tsinghua University, Beijing 100084, China}

\author{Jiu-Peng Chen}
\author{Yang Liu}
\affiliation{Hefei National Laboratory for Physical Sciences at Microscale and Department of Modern Physics, University of Science and Technology of China, Hefei 230026, China}
\affiliation{Shanghai Branch, CAS Center for Excellence and Synergetic Innovation Center in Quantum Information and Quantum Physics, University of Science and Technology of China, Shanghai 201315, China}
\affiliation{Jinan Institute of Quantum Technology, Jinan, Shandong 250101, China}

\author{Weijun Zhang}
\affiliation{State Key Laboratory of Functional Materials for Informatics, Shanghai Institute of Microsystem and Information Technology, Chinese Academy of Sciences, Shanghai 200050, China}

\author{Zong-Wen Yu}
\affiliation{State Key Laboratory of Low Dimensional Quantum Physics, Department of Physics, Tsinghua University, Beijing 100084, China}
\affiliation{Data Communication Science and Technology Research Institute, Beijing 100191, China}

\author{Hao Li}
\author{Lixing You}
\author{Zhen Wang}
\affiliation{State Key Laboratory of Functional Materials for Informatics, Shanghai Institute of Microsystem and Information Technology, Chinese Academy of Sciences, Shanghai 200050, China}

\author{Xiang-Bin Wang}
\affiliation{Shanghai Branch, CAS Center for Excellence and Synergetic Innovation Center in Quantum Information and Quantum Physics, University of Science and Technology of China, Shanghai 201315, China}
\affiliation{Jinan Institute of Quantum Technology, Jinan, Shandong 250101, China}
\affiliation{State Key Laboratory of Low Dimensional Quantum Physics, Department of Physics, Tsinghua University, Beijing 100084, China}

\author{Qiang Zhang}
\affiliation{Hefei National Laboratory for Physical Sciences at Microscale and Department of Modern Physics, University of Science and Technology of China, Hefei 230026, China}
\affiliation{Shanghai Branch, CAS Center for Excellence and Synergetic Innovation Center in Quantum Information and Quantum Physics, University of Science and Technology of China, Shanghai 201315, China}

\author{Jian-Wei Pan}
\affiliation{Hefei National Laboratory for Physical Sciences at Microscale and Department of Modern Physics, University of Science and Technology of China, Hefei 230026, China}
\affiliation{Shanghai Branch, CAS Center for Excellence and Synergetic Innovation Center in Quantum Information and Quantum Physics, University of Science and Technology of China, Shanghai 201315, China}

\clearpage
\begin{abstract}
Quantum key distribution can provide unconditionally secure key exchange for remote users in theory. In practice, however, in most quantum key distribution systems, quantum hackers might steal the secure keys by listening to the side channels in the source, such as the photon frequency spectrum, emission time, propagation direction, spatial angular momentum, and so on. It is hard to prevent such kinds of attacks because side channels may exist in any of the encoding space whether the designers take care of or not.
Here we report an experimental realization of a side-channel-free quantum key distribution protocol which is not only measurement-device-independent, but also immune to all side-channel attacks in the source.
We achieve a secure key rate of $4.80\times10^{-7}$ per pulse through 50 km fiber spools.

\end{abstract}

\maketitle

\section*{Introduction}
The cyber security today is protected by the modern cryptography, which is based on the computational complexity assumption. This assumption, however, might be challenged by the progress in algorithm~\cite{stevens2017first} or super computer~\cite{Shor1994algorithms,grover1996fast}.
Besides, hackers may steal the information from side channels instead of decrypting the ciphered message.
For example, one can attack the communication terminals that store the secret bits by detecting the physical effects like time shift~\cite{kocher1996timing}, power consumption~\cite{kocher1999differential}, electromagnetic leak~\cite{vaneck1985electromagnetic}, sound variation~\cite{genkin2014rsa} and etc.

Guaranteed by basic principles of quantum mechanics~\cite{wootters1982single}, quantum key distribution (QKD) generates information theoretically secure keys~\cite{bennett1984quantum,gisin2002quantum,hwang2003quantum,wang2005beating,lo2005decoy,scarani2009security,xu2020secure,pirandola2019advances} even if hackers have the most powerful attacks that physical laws permit.
However, side channels may appear in practical QKD systems due to device imperfections~\cite{xu2020secure}, leading to potential security loopholes. Actually, device imperfections, especially those in the detections, are the most serious threat to ``prepare-and-measure" QKD systems, such as time-shift attack~\cite{qi2007time,zhao2008quantum}, detector-blinding attack~\cite{makarov2009controlling,lydersen2010hacking}, detector-after-gate attack~\cite{wiechers2011after} and so on.
Luckily, this can be solved by measurement-device-independent QKD (MDIQKD)~\cite{lo2012measurement,braunstein2012side,tamaki2012phase,wang2013three,curty2014finite,xu2014protocol,zhou2016making,yin2016measurement}, which is immune to all attacks to measurement devices.
But the problem of side channels from the source still exists, leaving potential loopholes. Though the security is proven with the ideal encoding state, it can still be undermined when there is difference in the side channels of the emitted photons. For example, in a protocol using polarization encoding or phase encoding, there can be imperfections in side-channel space such as the frequency spectrum, the light emission time, etc.
These imperfections are highly possible, because the encodings in the source inevitably operate on a larger space. For example, the intensity modulation in the source may also affect the timing and frequency of the pulse. In such cases, the eavesdropper may acquire the secure keys by monitoring the side channels only, without affecting the encoding space. As a simple example, the eavesdropper may distinguish the intensity by monitoring the wavelength. Thus, the side channels actually undermine the security of practical QKD systems.

Recently, Wang et al proposed a novel side-channel-free (SCF) protocol~\cite{wang2019practical}.
This protocol is not only immune to all attacks in the side-channel space of emitted photons, such as the attacks on the imperfections of the frequency spectrum, emission time, non-ideal propagation direction, spatial angular momentum, etc, but also closes all potential loopholes in detection, by adapting the measurement-device-independent architecture, so it has a higher security level than most QKD protocols before. In the protocol, coherent state without any phase randomization is used as the source, so the decoy-state assumption is not required; Alice and Bob only decide on sending or not-sending the coherent state for encoding, so no more modulations are needed in the experiment. The theoretical proved side-channel-free characteristic from the source and the simple operation in encoding are the essential differences between the SCF-QKD\cite{wang2019practical} and the twin-field QKD (TF-QKD)~\cite{lucamarini2018overcoming}, including the sending-or-not-sending (SNS) protocol~\cite{wang2018twin}. We note that there are other protocols to achieve the side-channel-free security~\cite{braunstein2012side,mayers1998proceedings,acin2007device,scarani2008quantum}, but this protocol~\cite{wang2019practical} is the only one achieved with existing technologies, which is also possible to achieve a long distribution distance in practice.

Here, for the first time, we realize the side-channel-free QKD. Secure key rate of $4.80\times10^{-7}$ per pulse is achieved over 50 km.
Precise wavelength control and fast phase compensation have been utilized to accurately control and estimate the phase difference in the single-photon interference of two independent laser sources. High efficiency detection has also been applied to meet the demand of estimating phase drift between two users' fibers and improve the detection rates of signal pulses simultaneously.

\section*{Protocol}
To make the original protocol~\cite{wang2019practical} more practical, a revised side-channel-free protocol with phase reference pulses and phase postselection method is used which includes the following steps:

{\bf Step 1.} At each time window, Alice (Bob) prepares a nonrandom-phase coherent state
$\ket{\alpha_A} = e^{-\mu/2} \sum_{n=0}^\infty \frac{(\sqrt\mu e^{i\gamma_A})^n} {\sqrt{n!}} \ket{n}$ ($\ket{\alpha_B} = e^{-\mu/2} \sum_{n=0}^\infty \frac{(\sqrt\mu e^{i\gamma_B})^n} {\sqrt{n!}} \ket{n}$), where $\alpha_A=\sqrt\mu e^{i\gamma_A}$ ($\alpha_B=\sqrt\mu e^{i\gamma_B}$).
With probability $\varepsilon$, Alice (Bob) decides on {\em sending} and she (he) sends out the coherent state $|\alpha_A\rangle$ ($|\alpha_B\rangle$) to Charlie and puts down a classical bit value $1$ ($0$) locally; with probability $1-\varepsilon$, she (he) decides on {\em not-sending} and she (he) does not send out anything and puts down a classical bit value $0$ ($1$) locally. These pulses (coherent states or vacuum) are called signal pulses. She (He) also prepares a strong reference pulse time-multiplexed with the signal pulses. The phases of the reference pulses are modulated periodically and this will be presented in detail later. No matter what she (he) decides, she (he) always sends the reference pulse to Charlie.
They define a $\tilde{Z}$ window as a time window when either Alice or Bob decides on {\em sending} and the other decides on {\em not-sending}.

{\em Note:} Different from the decoy-state method requiring phase-randomized coherent states, here Alice and Bob are required to use the nonrandom-phase coherent states.
Their initial individual phases ($\gamma_A$ and $\gamma_B$) are fixed during the whole experiment. The reference pulse is introduced only to carry the information about the phase, which is allowed to be known by Eve according to the protocol. It has nothing to do with either the bit value or the state of the signal pulse (except the phase). Thus the introduction of reference pulse doesn't affect the security.

\begin{figure}[htb]
\centering
\resizebox{5cm}{!}{\includegraphics{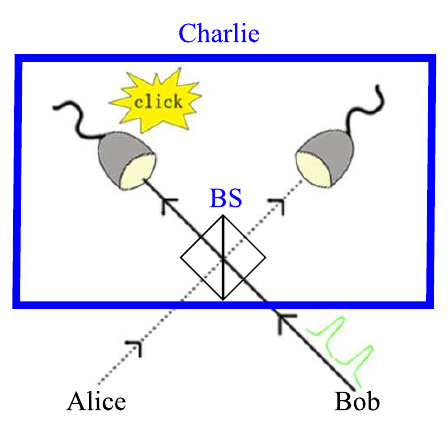}}
\caption{A schematic illustration of side-channel-free QKD. BS:beam-splitter.}
\label{Fig:schematicillustration}
\end{figure}

{\bf Step 2.} Charlie measures the signal pulses at the measurement station between Alice and Bob and announces which detector clicks. If one and only one detector clicks, this time window is regarded as an {\em effective time window}. In addition, he measures the reference pulses and announces the phase difference $\delta$ between Alice's and Bob's pulses to learn the phase shifts in the channels. It is shown schematically in Fig.~\ref{Fig:schematicillustration}.

{\bf Step 3.} According to Charlie's measurement results, Alice and Bob keep the bits from {\em all} events in which the phase difference $\delta$ satisfies the condition
\begin{equation}\label{equ:Condition}
  |\delta|<\Delta,
\end{equation}
where $\Delta$ is the relative phase difference threshold, and they announce the bit values of other bits, and then discard them.
Here $|x|$ means the degree of the minor angle enclosed by the two rays that enclose the rotational angle of degree $x$, e.g., $|-15\pi/8|=|15\pi/8|=\pi/8$.

{\bf Step 4.} Among the preserved bits, Alice and Bob take a random subset, $u$, through classical communication, to do error test and parameter estimation.
They announce all bit values in set $u$ through classical channel.
They discard the bits from the set $u$ after the error test, and the set of remaining bits is called the set $v$.

{\bf Step 5.} Alice and Bob distill (by conducting error correction and privacy amplification) the effective bits from the set $v$, with the asymptotic key rate for the number of final bits
\begin{equation}\label{equ:keylength}
    n_F = n_{\tilde Z}[1-H(\bar e^{ph})]- f n_v H(E_v)
\end{equation}
where $H(x)=-x\log_2 x -(1-x) \log_2 (1-x)$ is the entropy function; $n_{\tilde Z}$ ($n_v$) is the number of remaining effective bits from $\tilde Z$ windows (all time windows) in set $v$; $\bar e^{ph}$ is the upper bound of the phase-flip error rate for bits in effective $\tilde Z$ windows in set $v$; $f$ is the correction efficiency factor which we set $f=1.1$, and $E_v$ is the bit-flip error rate of effective bits in set $v$. The values of $n_{\tilde{Z}}$ and $\bar e^{ph}$ can be obtained by the observed data of the set $u$ asymptotically.

Equivalently, the key rate (per pulse) can be written as
\begin{equation}\label{equ:keyrate}
    R = \frac{1}{N_t} \{n_{\tilde Z}[1-H(\bar e^{ph})]- f n_v H(E_v)\},
\end{equation}
where $N_t$ is the total number of signal pulses that Alice (Bob) sends. The details of the calculation of the key rate are shown in the Supplemental Material.

\section*{Experiment}
\begin{figure*}[tbh]
\centering
\resizebox{14cm}{!}{\includegraphics{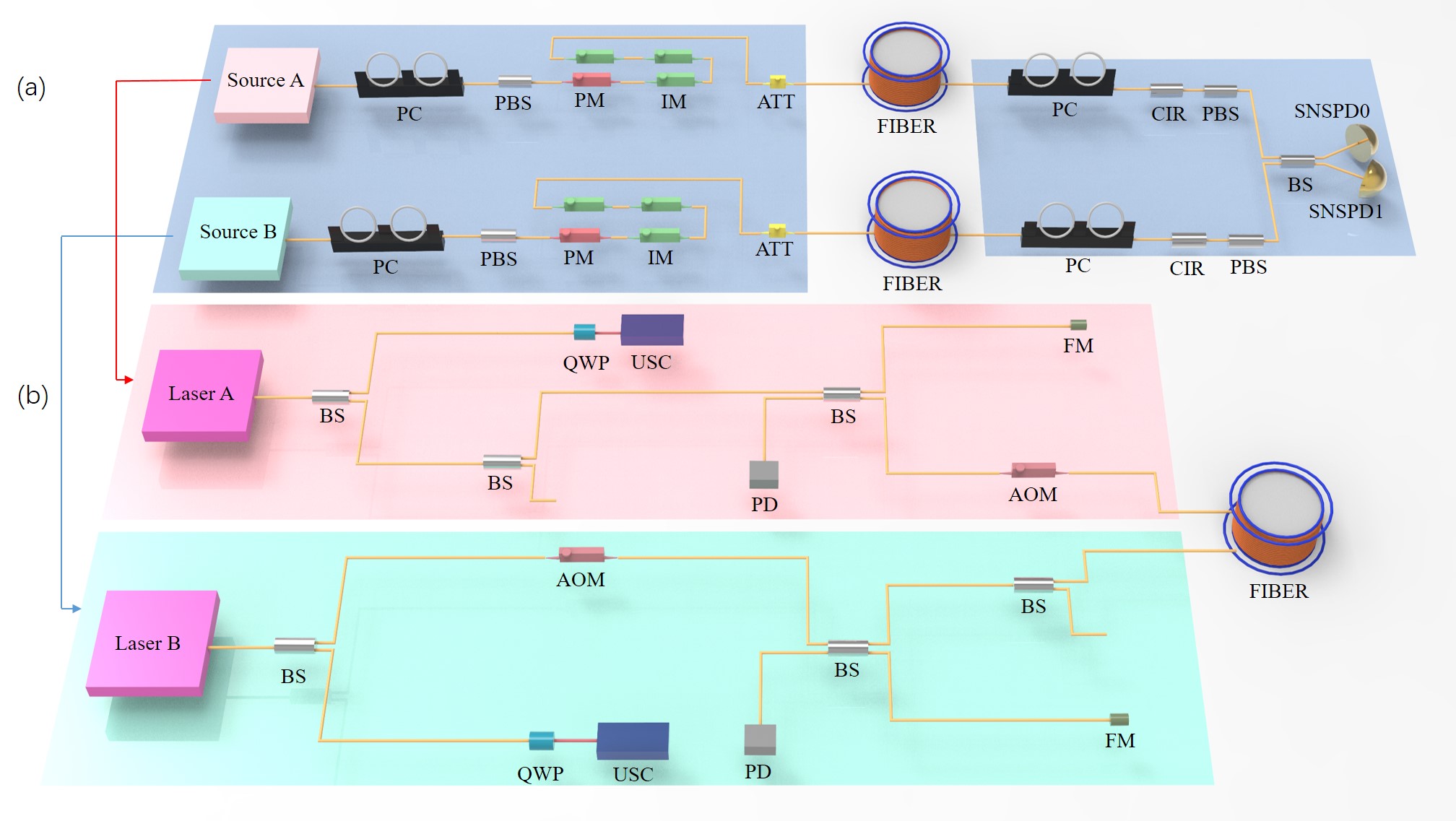}}
\caption{(a) Experimental setup. Alice and Bob modulate the phase-locked lasers with a phase modulator (PM) and three intensity modulators (IM1, IM2, IM3). The PM is used to encode the reference pulses; the intensity modulator IM1 is used to set ``sending" or ``not sending" of the signal, while IM2 and IM3 are used to set the intensities between the reference and signal pulses. Additional attenuators (ATTs) in the secure zones are calibrated to set the proper output photon intensity. Charlie interferes and measures the light from Alice and Bob with superconducting nanowire single-photon detectors (SNSPDs). PC: polarization controller, PBS: polarization beam splitter, CIR: circulator, BS: beam splitter. (b) Alice and Bob lock the wavelength of their lasers with a frequency-locking system~\cite{liu2019experimental}. The light transmits through an additional 50 km fiber. AOM: acousto-optic modulator; FM: Faraday mirror; PD: photodiode; QWP: quarter wave plate; USC: ultra-stable cavity.}
\label{Fig:setup}
\end{figure*}

To implement the side-channel-free protocol above, the experiment is designed as in Fig.~\ref{Fig:setup} (a). We use the time-frequency dissemination technology to accurately control the phase difference in the single-photon interference of two independent laser sources. Intensity modulators are used to control the ``sending" and ``not-sending" encoding. Finally, high performance superconducting nanowire single-photon detectors are used to meet the stringent peak counting rate requirement of the reference pulse detection, as well as high efficiency requirement of the signal pulses. In the following, we will discuss the experiment in detail.

First, stable lasers with exact wavelength are required in the remote single-photon interference in our SCF-QKD experiment.
The wavelengths of Alice's and Bob's independent lasers are locked with the time-frequency transmission technology~\cite{liu2019experimental}, through additional 50 km fiber spools shown in Fig.~\ref{Fig:setup} (b).
Alice uses a commercial sub-Hz laser source with a central wavelength of 1550.1665 nm and it is internally locked into her cavity; Bob uses a commercial kilo-Hz fiber laser, locked to an ultra-low-expansion(ULE) glass cavity with Pound-Drever-Hall (PDH) technique~\cite{drever1983laser,pound1946electronic}. The final linewidth is approximately 1 Hz with a central wavelength of 1550.1674 nm. Obviously the frequency difference of two laser sources about 112 Megahertz still exists. Therefore, at Bob's station, we insert an acoustic-optic modulator (AOM) with a tunable carrier frequency to compensate the frequency difference in real time. The phase noise via the 50 km fiber spools is cancelled in real time using another AOM with a carrier frequency of 40 Megahertz by Alice.

With these narrow linewidth coherent light sources prepared, the next step is to encode. As for signal pulses, only one intensity for ``sending" is required. We use an intensity modulator (IM) to modulate the signal to 1 ns pulse duration for ``sending" and to vacuum for ``not sending". Based on the experimental conditions, we optimize the coherent state intensity to $\mu=0.002$, and the probability of ``sending" to $\varepsilon=0.021$. The ``sending" and ``not sending" are determined by previously prepared quantum random numbers, with only one intensity modulation required.
While the relative phase between Alice and Bob is not stable: the fluctuation of the fiber length and refractive index directly affect the relative phase; the wavelength difference of the sources may also contribute to the phase drift.

In order to correct the relative phase drift, we adopt strong reference pulses to estimate the relative phase between Alice's and Bob's signal pulses~\cite{liu2019experimental}. For every 1 $\mu s$ time interval, 15 signal pulses are encoded in the first 450 ns as ``sending" or ``not sending"; then in the following 400 ns, 4 phase encoded reference pulses are sent; the final 150 ns are used as the recovery time for the superconducting nanowire single-photon detectors (SNSPDs), with vacuum states sent. Thus the effective signal frequency is 15 MHz considering the reference pulses. The reference intensity is set to $\mu_{ref}=0.062$ for each 1 ns width.

The relative intensities between signal and reference pulses are modulated with another two IMs. All three IMs are set to vacuum if the signal state is ``not sending" to increase the extinction ratio. To estimate the relative phases in the fibers, a phase modulator (PM) in Alice's station sets the phase of her reference pulses to \{$0$, $\pi$/2, $\pi$, 3$\pi$/2\} respectively, while all of Bob's phase reference pulses are set to $\pi$. Note that the phase modulation is only applied to Alice's and Bob's reference pulses to estimate relative phase drift in the fibers; the phases of the signals are always set to $0$.
In another word, the phase modulation works only as an estimation of the reference frame, thus it would not introduce side channels to the system.

\begin{figure}[htb]
\centering
\resizebox{8cm}{!}{\includegraphics{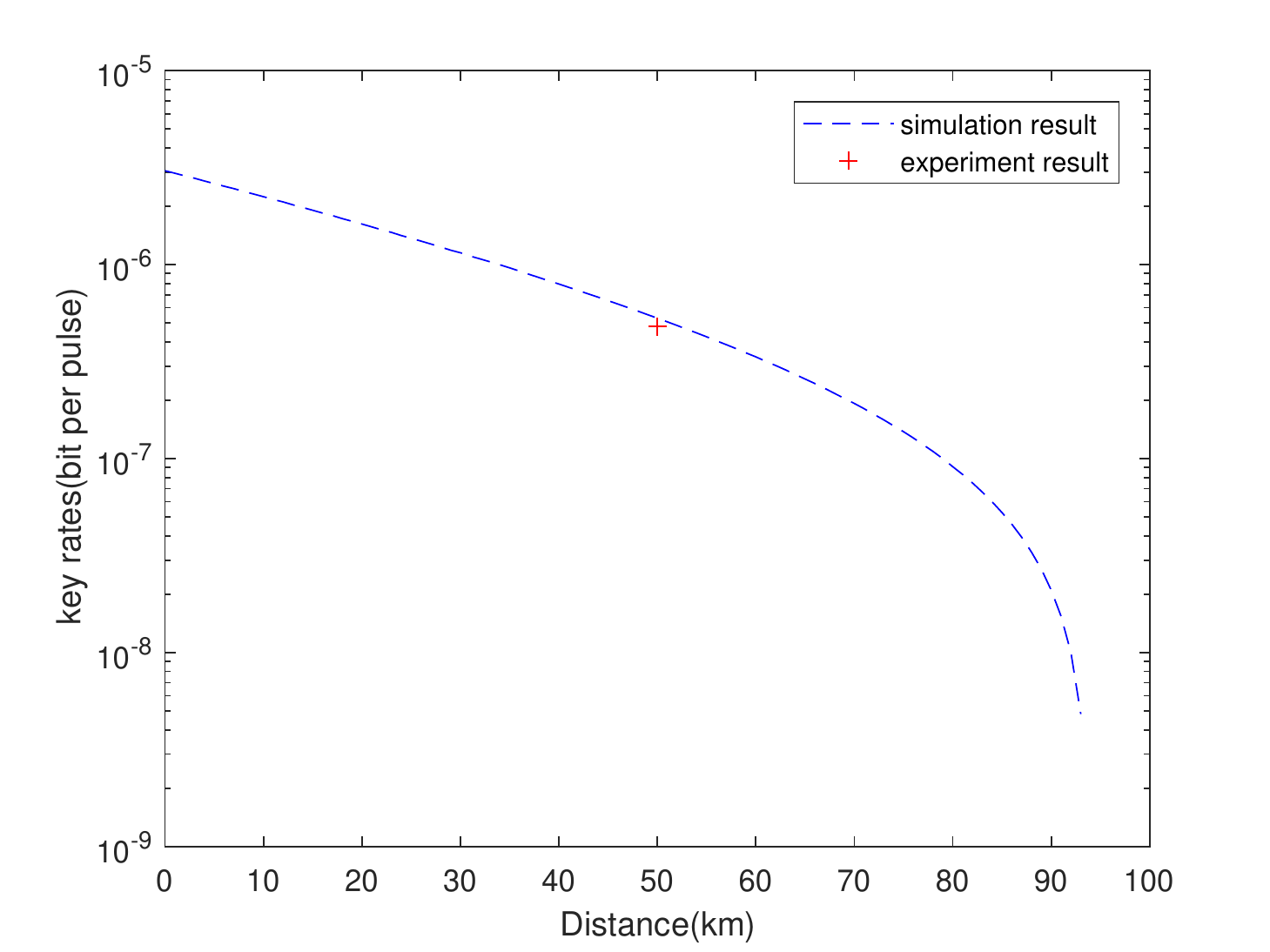}}
\caption{Secure key rates of the SCF-QKD. The dashed blue curve shows the simulated key rates in standard single mode fiber with the
coherent state intensity $\mu=0.002$, the ``sending" probability $\varepsilon=0.021$ and the phase difference threshold between Alice's and Bob's optical pulses $\Delta=30^\circ$. The red cross represents the experimental key rate over 50 km fiber spools. In both simulation and experiment, 10\% of the pulses after phase post-selection are used for error testing and parameter estimating.}
\label{Fig:keyrateVSdistance}
\end{figure}

The signals from Alice and Bob are transmitted through 25 km fiber spools respectively to the measurement station, Charlie, with the channel loss of 5.0 dB and 5.2 dB. The light is filtered with circulators to eliminate the SNSPD back scattered light. Then polarization controllers and polarization beam splitters are used to correct the polarization before interference at Charlie's beam splitter. The additional loss of the optical components are 4.78 dB for Alice and 4.44 dB for Bob. The interference results are measured with two SNSPDs with detection efficiencies  of 60.5\% and 62.6\%, and recorded by a time tagger. Here the SNSPDs are designed to achieve high detection efficiency and high peak counting rates of around 10 MHz simultaneously.

Following the phase estimation procedure~\cite{liu2019experimental} (See Supplemental Material for the details of phase estimation), Charlie calculates the relative phase based on the four reference pulses interference results due to the four corresponding encoded phases. For each effective detection event that one and only one detector clicks, the interference results of 48 nearest reference pulses within 12 $\mu$s are collected for estimation. On average, about 45 detections in this time period can be recorded for phase estimation which are enough for accurate phase estimation.

During a 12.0-hour experimental test, $6.04\times10^{11}$ signal pulses are sent, in which $7,004,417$ are detected and recorded. For each effective detection event, the relative phase between Alice's and Bob's signal pulses is calculated with the phase estimation procedure. Next, a threshold of relative phase difference $\Delta$ is set. Only the data with the relative phase $|\delta|<\Delta$ are kept as raw keys, as in Eq.~(\ref{equ:Condition}); for all other detections, Alice and Bob disclosed the bit values to calculate the state of the twin-field after phase postselection which is
\begin{equation}\label{equ:rho'}\begin{split}
		\rho^\prime=&c_1 \oprod{\alpha,\alpha}{\alpha,\alpha} + c_2 \oprod{\alpha,0}{\alpha,0}\\
		&+c_3 \oprod{0,\alpha}{0,\alpha} + c_4 \oprod{0,0}{0,0}
\end{split}\end{equation}
where $c_1=0.000442$, $c_2=0.020594$, $c_3=0.020564$, $c_4=0.958400$ according to the estimation (See Supplemental Material for details about the calculation).
After the phase postselection, a portion $p_{t}$ of the bits are selected as ``test bits". The values of the ``test bits" announced by Alice and Bob, as well as the detections, are then used for calculating the number of remaining effective bits in $\tilde{Z}$ window $n_{\tilde{Z}}$ and the upper bound of phase-flip error rate $\bar e^{ph}$ (See Supplemental Material for the theory and the details of parameters). The ``test bits" are then discarded.
In our experiment, the threshold of relative phase difference is set to $\Delta=30^\circ$ by optimization and the ``test bits" probability is set to $p_{t}=0.1$.
The number of remaining effective bits in $\tilde{Z}$ window and all time windows after the process of phase post-selection and ``test bits'' selection are $n_{\tilde{Z}}=2,207,341$ and $n_{v}=2,248,625$ respectively. The bit-flip error rate and the upper bound of phase-flip error rate are calculated to $E_v=2.12\%$ and $\bar e^{ph}=19.1\%$ respectively. Finally, we extracted 289,900 bit secure keys which is equivalent to $4.80\times10^{-7}$ per pulse. The key rate obtained in our experiment and the theoretical simulation are plotted in Fig.~\ref{Fig:keyrateVSdistance}. With our experimental parameters, it is predicted that a more than 80 km distribution distance can be achieved with our setup.

In conclusion, we have demonstrated the SCF-QKD protocol experimentally and obtained secure keys over 50 km fiber spools.
Our experiment shows that source-state side-channel-free and measurement-device-independent security can be simultaneously achieved in the QKD system with matured existing technologies.
The protocol makes no assumptions about the side-channel space of the quantum state though it assumes a perfect vacuum state in the encoding space for security proof. This assumption can possibly be loosened in future work.
We also note that instead of post-selecting phase difference between two fiber channels, active phase feedback can make more use of raw data and improve key rates of side-channel-free QKD. This shall be studied in the future.

\section*{Acknowledgments}
We would like to thank Feihu Xu for insightful discussions. This work was supported by the National Key R\&D Program of China (Grants No.2017YFA0303900, 2017YFA0304000, 2020YFA0309800), the National Natural Science Foundation of China, the Chinese Academy of Science (CAS), Shanghai Municipal Science and Technology Major Project (Grant No.2019SHZDZX01), Key R\&D Plan of Shandong Province (Grant No. 2019JZZY010205, 2020CXGC010105), the Shandong provincial natural science foundation (Grant No. ZR2020YQ45), the Taishan Scholar Program of Shandong Province, and Anhui Initiative in Quantum Information Technologies.

\clearpage
\begin{center}
\large
\textbf{Appendix}
\end{center}

\begin{appendix}
\section{Postselection instead of active phase compensation}
In the original protocol in \cite{wang2018twin}, Charlie is supposed to perform phase compensation to remove the phases of Alice's and Bob's coherent states, $\gamma_A$ and $\gamma_B$, and the phase shifts from the channels to obtain a low phase-flip error rate.
But this phase compensation is a bit difficult to perform in practice.
Instead, a revised protocol with phase reference pulses sent and postselection method can be used.
As described in {\bf Step 3} in Sec.\uppercase\expandafter{\romannumeral2} in the main text, Alice and Bob can perform postselection by only keeping the bits whose phase different $\delta$ satisfies the condition $|\delta|<\Delta$, using them to distill the final key, and discarding other bits.

Before postselection, the state of the twin-field sent by Alice and Bob is
\begin{equation}\begin{split}
		\rho=&\varepsilon^2 \oprod{\alpha,\alpha}{\alpha,\alpha} + \varepsilon(1-\varepsilon) \oprod{\alpha,0}{\alpha,0}\\
		&+(1-\varepsilon)\varepsilon \oprod{0,\alpha}{0,\alpha}+(1-\varepsilon)^2 \oprod{0,0}{0,0}
\end{split}\end{equation}
If the total number of pulse pairs sent is $N$, the numbers of these four twin-field states are $a_1=\varepsilon^2 N$, $a_2=\varepsilon(1-\varepsilon) N$, $a_3=\varepsilon(1-\varepsilon) N$, and $a_4=(1-\varepsilon)^2 N$, respectively.

According to the measurement results of calibration and $\delta$, Charlie announces which pulses are available and which are not.
Alice and Bob respectively announce what states they sent in those time windows with unavailable pulses, according to which they can calculate the new state of the twin-field after postselection.
Specifically, if the numbers of those four states in unavailable pulses are $b_1$, $b_2$, $b_3$, and $b_4$, respectively, the number of those available pulses is $N^\prime=N-b_1-b_2-b_3-b_4$ and the state of them is
\begin{equation}\label{equ:rho'}\begin{split}
		\rho^\prime=&c_1 \oprod{\alpha,\alpha}{\alpha,\alpha} + c_2 \oprod{\alpha,0}{\alpha,0}\\
		&+c_3 \oprod{0,\alpha}{0,\alpha} + c_4 \oprod{0,0}{0,0}
\end{split}\end{equation}
where $c_i=(a_i-b_i)/N^\prime,i=1,2,3,4$.
Then they use effective events from these $N^\prime$ events to do parameter estimation and distill the final key.
Note that Charlie may not be honest about the measurement results, but it doesn't affect the security of this revised protocol because the parameter estimation and key distillation are done by Alice and Bob themselves.

\section{Calculation of the key rate}
The set of events after postselection is defined as set $V$.
Among these events, Alice and Bob randomly choose $p_t=10\%$ from them to obtain the set $u$, which is used for error test and parameter estimation.
The set of remaining events after postselection is defined as set $v$, i.e. $u\cup v=V$.
They announce all bit values in set $u$ through classical channel, and obtain the numbers of each twin-field state that was sent, denoted as $N_{ab,u}$, the numbers of effective events from each state, denoted as $n_{ab,u}$, and the numbers of effective events causing the detector $i$ clicking, denoted as $n_{ab,u}^{i}$ with $i=L,R$.
In the subscript,   $a$ ($b$) can be 0 or $\alpha$, corresponding to Alice's (Bob's) decision of not-sending or sending, respectively, and ``$u$''means the number in set $u$.
For example, $n_{\alpha0,u}^R$ denotes the number of effective events in set $u$ when Alice decides to send and Bob decides not to send, and the right detector clicks.
Similarly, they can define $N_{ab,v}$ as the numbers of each twin-field state that was sent in set $v$ and calculate these values through Eq.~(\ref{equ:rho'}) and the value of $p_t$.

With the observed values above, they can calculate the counting rate of each twin-field state
\begin{equation}\label{equ:counting_rate}
	S_{ab,u}=n_{ab,u}/N_{ab,u},
\end{equation}
the  counting rate of each twin-field state with detector $i$ clicking
\begin{equation}\label{equ:counting_rate_det}
	S_{ab,u}^{i}=n_{ab,u}^{i}/N_{ab,u},
\end{equation}
the total counting rate of set $u$
\begin{equation}\label{equ:counting_rate_u}
	S_{u}=\sum_{a,b}n_{ab,u}/\sum_{a,b}N_{ab,u},
\end{equation}
and the bit-flip error rate of set $u$
\begin{equation}\label{equ:bit_error_u}
	E_{u}=(n_{00,u}+n_{\alpha\alpha,u})/\sum_{a,b}n_{ab,u},
\end{equation}

According to the formulas in Ref.~\cite{wang2019practical}, they can calculate the counting rate and the phase-flip error rate of $\tilde{Z}$ windows through the data of set $u$ asymptotically.
Specifically, in the asymptotic case, the counting rate of $\tilde{Z}$ windows is
\begin{equation}\label{equ:StildeZ}
	S_{\tilde{Z}}=\frac{1}{2}(S_{0\alpha,u}+S_{\alpha0,u}).
\end{equation}
Thus, the number of $\tilde{Z}$ windows in set $v$ is
\begin{equation}\label{equ:ntildeZ}
	n_{\tilde{Z}}=2\min(N_{0\alpha,v},N_{\alpha0,v})S_{\tilde{Z}}.
\end{equation}
Note that as the security proof in Ref.~\cite{wang2019practical} requires, the state in $\tilde{Z}$ windows has to be $\rho_{\tilde{Z}}=(\oprod{0,\alpha}{0,\alpha}+\oprod{\alpha,0}{\alpha,0})/2$, thus we use $\min(N_{0\alpha,v},N_{\alpha0,v})$ in Eq.~(\ref{equ:ntildeZ}).
And the upper bound of the phase-flip error rate of $\tilde{Z}$ windows is~\cite{wang2019practical}
\begin{equation}\label{equ:phasef}
	e^{ph}\le \overline{e}^{ph} = \frac{(1+e^{-\mu}) \left[\overline{S}_{X_+}^{R}-\underline{S}_{X_+}^{L}\right] +S_{0\alpha,u}^{L}+S_{\alpha0,u}^{L}} {2S_{\tilde Z}}
\end{equation}
where
\begin{equation}\label{equ:up}
	\begin{split}
		\overline{S}_{X_+}^{R} = \frac{1}{2(1+e^{-\mu})} \{e^{-\mu} S_{00,u}^{R} + \frac{1}{e^{-\mu}} S_{\alpha\alpha,u}^{R} + \frac{(1-e^{-\mu})^2}{e^{-\mu}} \\
		+  2\sqrt{S_{00,u}^{R} S_{\alpha\alpha,u}^{R}} + 2(1-e^{-\mu}) \sqrt{S_{00,u}^{R}} + \frac{2(1-e^{-\mu})}{e^{-\mu}} \sqrt{S_{\alpha\alpha,u}^{R}} \}
	\end{split}
\end{equation}
and
\begin{equation}\label{equ:down}
	\begin{split}
		\underline{S}_{X_+}^{L} = \frac{1}{2(1+e^{-\mu})} \bigg\{e^{-\mu} S_{00,u}^{L} + \frac{1}{e^{-\mu}} S_{\alpha\alpha,u}^{L} - \big[2\sqrt{S_{00,u}^{L} S_{\alpha\alpha,u}^{L}} \\+ 2(1-e^{-\mu}) \sqrt{S_{00,u}^{L}}
		 + \frac{2(1-e^{-\mu})}{e^{-\mu}} \sqrt{S_{\alpha\alpha,u}^{L}}\big] \bigg\}.
	\end{split}
\end{equation}

Then they can calculate the final key rate with Eq. (3) in the main text.

\section{Signal modulation}
Alice and Bob use independent continuous wave (CW) lasers as their light sources, with their frequencies locked to each other. They use 2 GHz sampling rate, 14-bit depth AWGs to modulate the CW lasers to generate the signal and reference pulses. The signal pattern is created with $3.02\times 10^{6}$ pre-generated quantum random numbers, and loaded to the AWGs prior to the experiment. Then a start signal from a controlling computer triggers the AWGs to send signals at the same time. The maximum amplitude generated by the AWGs is 500 mV, which is then amplified by about 25 dB for the electro-optical modulators.

\begin{figure}[htb]
\centering
\resizebox{8cm}{!}{\includegraphics{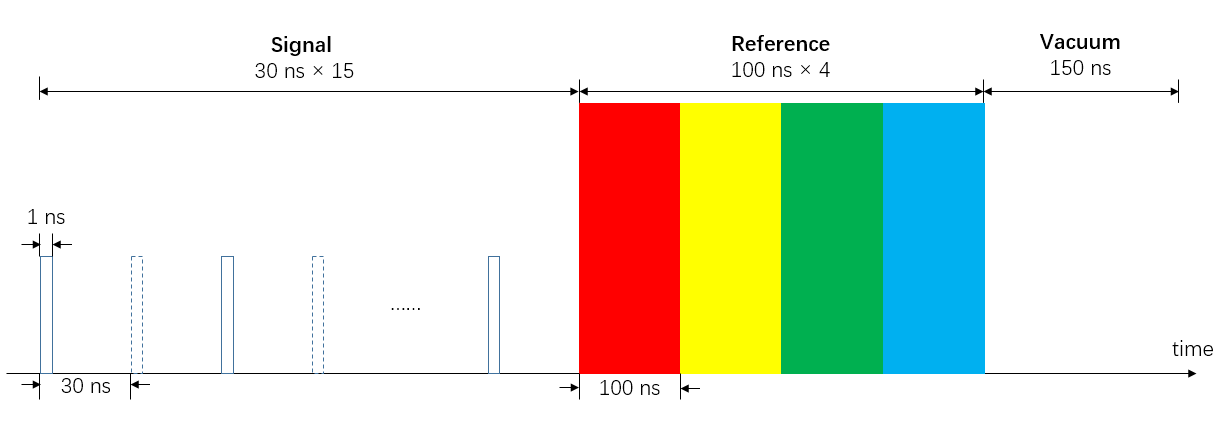}}
\caption{Alice's encoding pattern.}
\label{Fig:AliceEncodingPattern}
\end{figure}

\begin{figure}[htb]
\centering
\resizebox{8cm}{!}{\includegraphics{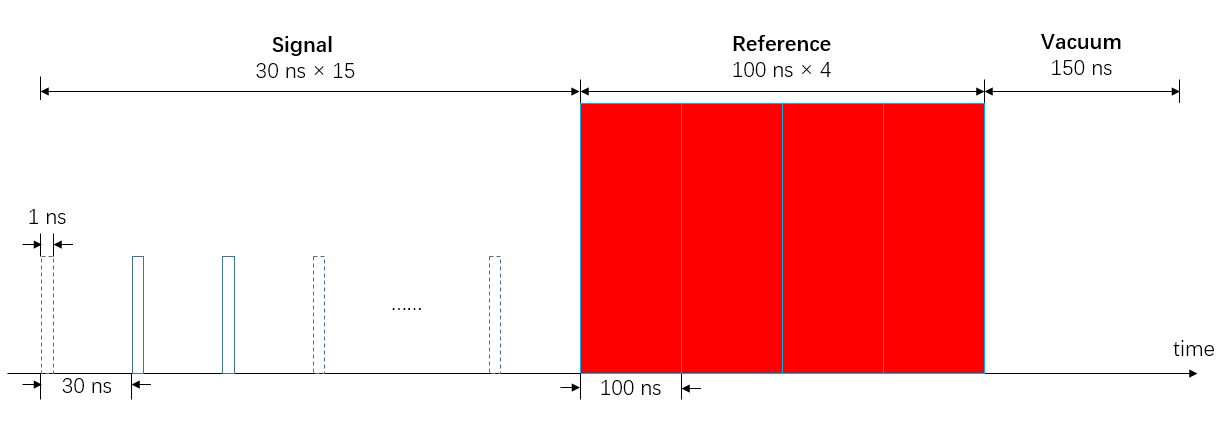}}
\caption{Bob's encoding pattern.}
\label{Fig:BobEncodingPattern}
\end{figure}

In the encoding, the signal and reference pulses are sent in each 1 $\mu$s time interval. As shown in Fig.~\ref{Fig:AliceEncodingPattern} and Fig.~\ref{Fig:BobEncodingPattern}, 15 signal pulses are sent in the first 450 ns, with 1 ns pulse duration and 30 ns period. Each signal pulse is either set to $\mu=0.002$ for bit ``1'', or completely blocked for bit ``0'', determined by the random signals from the AWGs. In the next 400 ns, 4 phase reference pulses, with 100 ns pulse width, are sent for estimating the relative phase between Alice's and Bob's fiber channels. Finally, a 150 ns vacuum state is set, waiting the SNSPDs to recover from high counting region of reference pulses.

For the modulation, a phase modulator (PM) followed by three intensity modulators (IMs) are used to generate the above mentioned signal pattern. The phase modulator is set before any intensity modulator so the signal phase will not affected by the wavelength dependent behavior of the phase modulator. The phase modulator is used to modulate the phase reference pulses only, setting Alice's phase reference to \{0, $\pi/2$, $\pi$, $3\pi/2$\}, and Bob's all four phase references to $\pi$. The phases are set to $0$ for all the signal pulses and in the 150 ns recovery time.

For the intensity modulation, IM1 is used to modulate the light to 1 ns pulse width for the ``sending'' bit, or to vacuum state for ``not-sending'' bit; IM2 and IM3 are used to adjust the intensities between the phase reference and the signal pulses. All the modulators block the light when a vacuum state needs to be sent. The signals are further attenuated to the single-photon level by an attenuator.

\section{Controlling the Relative Phase between Alice and Bob}
Due to the single-photon interference requirement, it is essential to compensate the phase difference between Alice and Bob. The phase difference originates either from the wavelength difference of the two lasers, or from the phase drift in the fiber channels. As discussed in the main text and in~\cite{liu2019experimental}, we lock the wavelength of the light sources in real time with time-frequency-dissemination technology.
The phase reference pulses and their interference results are used to estimate or compensate the relative phase drift between fibers.

Besides the phase modulation mentioned above, we set the reference pulses intensity to around 2 MHz counts for each superconducting nanowire single-photon detector (SNSPD) at Charlie's station, so that we have enough data for estimation and it is not too high to affect the SNSPD performance. Charlie counts the total detections of the two SNSPDs of each phase difference during this time as $N_{i}$, where $i = \{1,2,3,4\}$ represents the modulated phase difference between Alice and Bob of \{$\pi$, $3\pi/2$, 0, $\pi/2$\}. Then he calculates the probability:
\begin{equation}
p_{i}=2 N_{i}/\Sigma N_{i}
\label{Eq:Prop}
\end{equation}
and the theoretical probabilities with relative phase induced by the fiber channel $\Delta\varphi_T$:
\begin{equation}
p_{Ti}(\Delta\varphi_T)=\cos^2(\frac{\Delta\theta_i+\Delta\varphi_T}{2})
\end{equation}
where the phase differences are $\Delta\theta_i=$\{$0$, $\pi/2$, $\pi$, $3\pi/2$\}, for $i=\{1,2,3,4\}$ accordingly.

Then he minimizes the error $Err(\Delta\varphi_T)$ between measured and theoretical probabilities to estimate $\Delta\varphi_T$:
\begin{equation}
Err(\Delta\varphi_T) = \sum_i{[p_i-p_{Ti}(\Delta\varphi_T)]^2}
\label{Eq:Err}
\end{equation}

We set the phase reference statistical time to 12 $\mu$s. On average, 45 counts can be collected for the 4 reference pulses. During this time the relative phase in fiber may drift about 0.073 rad (or 4.2 degrees) which may induce less than 3.5\% error rate to the system when interfering.

With this phase estimation, we revise the SCF-QKD protocol as mentioned in the main text, to compensate the phase drifts via post-processing. The phase slice criterion is used to post-select signals. The phase slice is optimized to yield a maximum key rate~\cite{liu2019experimental}:
\begin{equation}\label{equ:condition}
  |\delta|<\Delta,
\end{equation}
where $\delta$ is the estimated values of phase difference for each signal pulse pair, and $|x|$ means the degree of the minor angle enclosed by the two rays that enclose the rotational angle of degree $x$, e.g., $|-15\pi/8|=|15\pi/8|=\pi/8$. And $\Delta$ is the relative phase difference threshold which is optimized to $30^\circ$ in the experiment. Note that active phase feedback that compensates the phase difference $\delta$ to around 0 in real time is also possible. This method shall be studied in the future.

\section{Experimental Parameters}
The main parameters used in our experiment are summarized in Tab.~\ref{Tab:Parameters}. The fiber length between Alice and Charlie and that between Bob and Charlie are set to the same, the ``Fiber Length" in the table is the total fiber length. $\mu$ is the intensity of the coherent state of signal pulse. The signal pulse width is 1 ns and the phase reference pulse width is 100 ns; as to compare, $\mu_{ref}$ is the reference intensity in 1 ns width. The ``sending" probability is denoted by $p_{S}$. And we calculated the ``Output Intensity" at Alice's (Bob's) output.
Finally, we listed our QKD system frequency, the equivalent system frequency (considering phase reference pulses) and the working hours in the experiment.

\begin{table}[htb]
\centering
  \caption{Experimental parameters.}
\begin{tabular}{ccccc}
\hline
Fiber Length & 50 km\\
\hline
$\mu$		& 0.002\\
$\mu_{ref}$	& 0.062\\
\hline
$p_{S}$		& 0.021\\
Output Intensity (pW)		& 3.184\\
\hline
System Frequency (MHz)             & 33.3\\
Equivalent System Frequency (MHz)  & 15.0\\
Working Hours                      & 12.0\\
\hline
\end{tabular}
\label{Tab:Parameters}
\end{table}

\section{Experimental Results}

We summarized the fiber and the optical elements transmittance and the SNSPD efficiencies in Tab.~\ref{Tab:Characterization}. The optical elements include the polarization controllers (PCs), the circulators (CIRs), the polarization beam splitters (PBSs), and the beam splitter (BS). The results are given for each of the two inputs (A/B) and outputs (ch0/ch1) as appropriate. The SNSPD efficiency includes the PC efficiency.

The experimental results are summarized in Tab.~\ref{Tab:Result}, including the final key rate $R$, the total signal pulses sent in the experiment $N_{total}$, the number of remaining effective bits $n_{\tilde{Z}}$ from $\tilde Z$ windows after the error test, the counting rate $s$, the upper bound of the phase error rate $\bar e^{ph}$ and the bit-flip error rate $E_v$ of set $v$, with the optimized accepted phase difference threshold $\Delta$ (in degrees). For error testing and parameter estimating, a fraction of signal pulses are sampled as ``test bits" after phase postselection with a portion of $p_{t}$. A digital gate is applied to select only the central part of the signal pulse, to avoid imperfect interference due to timing difference. The width of the gate is labelled as $t_{gate}$ and the fraction of signal in the gate is denoted as $r_{gate}$.

\begin{table}[htb]
\centering
  \caption{Efficiencies of fibers and measurement station.}
\begin{tabular}{ccccc}
\hline
$\eta_{FiberA}$	& 0.316\\
$\eta_{FiberB}$	& 0.310\\
\hline
PC-A & \multicolumn{4}{c}{95.4\%}\\
PC-B & \multicolumn{4}{c}{96.7\%}\\
\hline
CIR-A & \multicolumn{4}{c}{83.4\%}\\
CIR-B & \multicolumn{4}{c}{87.0\%}\\
\hline
PBS-A & \multicolumn{4}{c}{93.4\%}\\
PBS-B & \multicolumn{4}{c}{95.6\%}\\
\hline
BS-A-ch0 & \multicolumn{4}{c}{45.3\%}\\
BS-A-ch1 & \multicolumn{4}{c}{43.5\%}\\
BS-B-ch0 & \multicolumn{4}{c}{45.5\%}\\
BS-B-ch1 & \multicolumn{4}{c}{44.8\%}\\
\hline
SNSPD-ch0 & \multicolumn{4}{c}{60.52\%}\\
SNSPD-ch1 & \multicolumn{4}{c}{62.61\%}\\
\hline
\end{tabular}
\label{Tab:Characterization}
\end{table}

We summarized the raw data used for the calculations in Tab.~\ref{Tab:Result1}.
The number of pulses Alice and Bob send is labelled as ``Sent-CD", where ``C" (``D") is ``1" or ``0", indicating the intensity Alice (Bob) for ``sending'' or ``not-sending''. The number of pulses falling within the accepted phase difference range $\Delta$ is listed as ``Sent-CD-$\Delta$". The number of signal pulses falling within the accepted range $\Delta$ is listed as ``Sent-ABCD-$\Delta$", where ``A" (``B") is ``S" or ``T" indicating signal mode or test mode from Alice (Bob); the number of detections is listed as ``Detected-ABCD-Ch" which means the detections falling within the accepted difference range $\Delta$, where ``Ch" indicates the detection channel.
Note that the correct bits are the effective detections when Alice decides on ``sending" and Bob decides on ``not-sending", or vice versa.

\begin{table*}[htb]
\centering
  \caption{Experimental results}
\begin{tabular}{ccccc}
\hline
Fiber Length & 50 km\\
$R$			& $4.799\times 10^{-7}$ \\
$N_{total}$	& \multicolumn{4}{c}{$6.0396\times 10^{11}$}\\
$n_{\tilde{Z}}$   & $2.21\times10^6$ \\
$s$		& $2.77\times 10^{-4}$ \\
$\bar e^{ph}$	& 19.1\% \\
$E_v$ 	& 2.12\% \\
\hline
$p_{t}$       & 0.1 \\
$\Delta$		& $30^\circ$ \\
Signal Pulse width (ns) & 1 \\
$t_{gate}$ (ns)	& 0.85 \\
$r_{gate}$	& 0.67 \\
\hline
\end{tabular}
\label{Tab:Result}
\end{table*}

\begin{table*}[htb]
\centering
  \caption{Raw data}
\begin{tabular}{cccc}
\hline
Sent-00	& 578835000000 &Detected-SS00-ch0	& 1285\\
Sent-01	& 12438400000  &Detected-SS00-ch1	& 1220\\
Sent-10	& 12420000000  &Detected-SS01-ch0	& 546658\\
Sent-11	& 266800000  &Detected-SS01-ch1	& 545018\\
Sent-00-$\Delta$	& 206492000000  &Detected-SS10-ch0	& 554790\\
Sent-01-$\Delta$	& 4436985713  &Detected-SS10-ch1	& 554207\\
Sent-10-$\Delta$	& 4430718614  &Detected-SS11-ch0	& 42800\\
Sent-11-$\Delta$	& 95191234  &Detected-SS11-ch1	& 2647\\
Sent-SS00-$\Delta$	& 185843000000  &Detected-TT00-ch0	& 149\\
Sent-SS01-$\Delta$	& 3993295035  &Detected-TT00-ch1	& 126\\
Sent-SS10-$\Delta$	& 3987675420  &Detected-TT01-ch0	& 60690\\
Sent-SS11-$\Delta$	& 85674748  &Detected-TT01-ch1	& 60926\\
Sent-TT00-$\Delta$	& 20649175977  &Detected-TT10-ch0	& 62121\\
Sent-TT01-$\Delta$	& 443690678  &Detected-TT10-ch1	& 61683\\
Sent-TT10-$\Delta$	& 443043194  &Detected-TT11-ch0	& 4728\\
Sent-TT11-$\Delta$	& 9516486  &Detected-TT11-ch1	& 315\\
\hline
\end{tabular}
\label{Tab:Result1}
\end{table*}

Additionally, we also calculated the $QBERs$ when Alice and Bob both sent coherent states with different detection counts according to different phase difference thresholds $\Delta$. Finally we extracted the optimized secure key rates with the parameter values. These are all listed in Tables.~\ref{Tab:ResultQBER11}.

\begin{table*}[htb]
\centering
  \caption{QBERs and Detections for both Alice and Bob sent coherent states and Key Rate for different $\Delta$.}
\begin{tabular}{cccccccccc}
\hline
Results$\mid$$\Delta$ & $2^\circ$	& $5^\circ$	& $8^\circ$	& $10^\circ$	& $12^\circ$	& $15^\circ$	& $30^\circ$	& $45^\circ$ \\
\hline
QBERs	& 3.4\% & 3.4\% & 3.6\% & 3.8\% & 4.0\% & 4.1\% & 5.9\% & 8.5\%  \\
Detections	& 4401 & 9169 & 14223 & 17522 & 20937 & 25981 & 50490 & 74728  \\
Key rates   &$7.19\times 10^{-8}$   & $1.26\times 10^{-7}$   & $2.29\times 10^{-7}$  & $2.26\times 10^{-7}$   & $3.29\times 10^{-7}$    & $2.95\times 10^{-7}$    &$4.80\times 10^{-7}$  & $4.73\times 10^{-7}$  \\
\hline
\end{tabular}
\label{Tab:ResultQBER11}
\end{table*}

\end{appendix}
\clearpage

\bibliographystyle{apsrev}
\bibliography{SCF_Exp_arXiv}

\begin{thebibliography}{38}
\expandafter\ifx\csname natexlab\endcsname\relax\def\natexlab#1{#1}\fi
\expandafter\ifx\csname bibnamefont\endcsname\relax
  \def\bibnamefont#1{#1}\fi
\expandafter\ifx\csname bibfnamefont\endcsname\relax
  \def\bibfnamefont#1{#1}\fi
\expandafter\ifx\csname citenamefont\endcsname\relax
  \def\citenamefont#1{#1}\fi
\expandafter\ifx\csname url\endcsname\relax
  \def\url#1{\texttt{#1}}\fi
\expandafter\ifx\csname urlprefix\endcsname\relax\def\urlprefix{URL }\fi
\providecommand{\bibinfo}[2]{#2}
\providecommand{\eprint}[2][]{\url{#2}}

\bibitem[{\citenamefont{Stevens et~al.}(2017)\citenamefont{Stevens, Bursztein,
  Karpman, Albertini, and Markov}}]{stevens2017first}
\bibinfo{author}{\bibfnamefont{M.}~\bibnamefont{Stevens}},
  \bibinfo{author}{\bibfnamefont{E.}~\bibnamefont{Bursztein}},
  \bibinfo{author}{\bibfnamefont{P.}~\bibnamefont{Karpman}},
  \bibinfo{author}{\bibfnamefont{A.}~\bibnamefont{Albertini}},
  \bibnamefont{and} \bibinfo{author}{\bibfnamefont{Y.}~\bibnamefont{Markov}},
  in \emph{\bibinfo{booktitle}{Advances in Cryptology -- CRYPTO 2017}}
  (\bibinfo{year}{2017}), pp. \bibinfo{pages}{570--596}.

\bibitem[{\citenamefont{{Shor}}(1994)}]{Shor1994algorithms}
\bibinfo{author}{\bibfnamefont{P.~W.} \bibnamefont{{Shor}}}, in
  \emph{\bibinfo{booktitle}{Proceedings 35th Annual Symposium on Foundations of
  Computer Science}} (\bibinfo{year}{1994}), pp. \bibinfo{pages}{124--134}.

\bibitem[{\citenamefont{Grover}(1996)}]{grover1996fast}
\bibinfo{author}{\bibfnamefont{L.~K.} \bibnamefont{Grover}}, in
  \emph{\bibinfo{booktitle}{Proceedings of the Twenty-Eighth Annual ACM
  Symposium on Theory of Computing}} (\bibinfo{year}{1996}), pp.
  \bibinfo{pages}{212--219}.

\bibitem[{\citenamefont{Kocher}(1996)}]{kocher1996timing}
\bibinfo{author}{\bibfnamefont{P.~C.} \bibnamefont{Kocher}}, in
  \emph{\bibinfo{booktitle}{Advances in Cryptology --- CRYPTO '96}}, edited by
  \bibinfo{editor}{\bibfnamefont{N.}~\bibnamefont{Koblitz}}
  (\bibinfo{year}{1996}), pp. \bibinfo{pages}{104--113}.

\bibitem[{\citenamefont{Kocher et~al.}(1999)\citenamefont{Kocher, Jaffe, and
  Jun}}]{kocher1999differential}
\bibinfo{author}{\bibfnamefont{P.}~\bibnamefont{Kocher}},
  \bibinfo{author}{\bibfnamefont{J.}~\bibnamefont{Jaffe}}, \bibnamefont{and}
  \bibinfo{author}{\bibfnamefont{B.}~\bibnamefont{Jun}}, in
  \emph{\bibinfo{booktitle}{Advances in Cryptology --- CRYPTO' 99}}, edited by
  \bibinfo{editor}{\bibfnamefont{M.}~\bibnamefont{Wiener}}
  (\bibinfo{year}{1999}), pp. \bibinfo{pages}{388--397}.

\bibitem[{\citenamefont{{van Eck}}(1985)}]{vaneck1985electromagnetic}
\bibinfo{author}{\bibfnamefont{W.}~\bibnamefont{{van Eck}}},
  \bibinfo{journal}{Computers \& Security} \textbf{\bibinfo{volume}{4}},
  \bibinfo{pages}{269} (\bibinfo{year}{1985}).

\bibitem[{\citenamefont{Genkin et~al.}(2014)\citenamefont{Genkin, Shamir, and
  Tromer}}]{genkin2014rsa}
\bibinfo{author}{\bibfnamefont{D.}~\bibnamefont{Genkin}},
  \bibinfo{author}{\bibfnamefont{A.}~\bibnamefont{Shamir}}, \bibnamefont{and}
  \bibinfo{author}{\bibfnamefont{E.}~\bibnamefont{Tromer}}, in
  \emph{\bibinfo{booktitle}{Advances in Cryptology -- CRYPTO 2014}}, edited by
  \bibinfo{editor}{\bibfnamefont{J.~A.} \bibnamefont{Garay}} \bibnamefont{and}
  \bibinfo{editor}{\bibfnamefont{R.}~\bibnamefont{Gennaro}}
  (\bibinfo{year}{2014}), pp. \bibinfo{pages}{444--461}.

\bibitem[{\citenamefont{Wootters and Zurek}(1982)}]{wootters1982single}
\bibinfo{author}{\bibfnamefont{W.}~\bibnamefont{Wootters}} \bibnamefont{and}
  \bibinfo{author}{\bibfnamefont{W.}~\bibnamefont{Zurek}},
  \bibinfo{journal}{Nature} \textbf{\bibinfo{volume}{299}},
  \bibinfo{pages}{802} (\bibinfo{year}{1982}).

\bibitem[{\citenamefont{BENNETT}(1984)}]{bennett1984quantum}
\bibinfo{author}{\bibfnamefont{C.}~\bibnamefont{BENNETT}}, in
  \emph{\bibinfo{booktitle}{Proceedings of the IEEE International\ Conference
  on Computers, Systems, and Signal Processing}} (\bibinfo{year}{1984}), pp.
  \bibinfo{pages}{175--179}.

\bibitem[{\citenamefont{Gisin et~al.}(2002)\citenamefont{Gisin, Ribordy,
  Tittel, and Zbinden}}]{gisin2002quantum}
\bibinfo{author}{\bibfnamefont{N.}~\bibnamefont{Gisin}},
  \bibinfo{author}{\bibfnamefont{G.}~\bibnamefont{Ribordy}},
  \bibinfo{author}{\bibfnamefont{W.}~\bibnamefont{Tittel}}, \bibnamefont{and}
  \bibinfo{author}{\bibfnamefont{H.}~\bibnamefont{Zbinden}},
  \bibinfo{journal}{Reviews of modern physics} \textbf{\bibinfo{volume}{74}},
  \bibinfo{pages}{145} (\bibinfo{year}{2002}).

\bibitem[{\citenamefont{Hwang}(2003)}]{hwang2003quantum}
\bibinfo{author}{\bibfnamefont{W.-Y.} \bibnamefont{Hwang}},
  \bibinfo{journal}{Physical Review Letters} \textbf{\bibinfo{volume}{91}},
  \bibinfo{pages}{057901} (\bibinfo{year}{2003}).

\bibitem[{\citenamefont{Wang}(2005)}]{wang2005beating}
\bibinfo{author}{\bibfnamefont{X.-B.} \bibnamefont{Wang}},
  \bibinfo{journal}{Physical Review Letters} \textbf{\bibinfo{volume}{94}},
  \bibinfo{pages}{230503} (\bibinfo{year}{2005}).

\bibitem[{\citenamefont{Lo et~al.}(2005)\citenamefont{Lo, Ma, and
  Chen}}]{lo2005decoy}
\bibinfo{author}{\bibfnamefont{H.-K.} \bibnamefont{Lo}},
  \bibinfo{author}{\bibfnamefont{X.}~\bibnamefont{Ma}}, \bibnamefont{and}
  \bibinfo{author}{\bibfnamefont{K.}~\bibnamefont{Chen}},
  \bibinfo{journal}{Physical Review Letters} \textbf{\bibinfo{volume}{94}},
  \bibinfo{pages}{230504} (\bibinfo{year}{2005}).

\bibitem[{\citenamefont{Scarani et~al.}(2009)\citenamefont{Scarani,
  Bechmann-Pasquinucci, Cerf, Du{\v{s}}ek, L{\"u}tkenhaus, and
  Peev}}]{scarani2009security}
\bibinfo{author}{\bibfnamefont{V.}~\bibnamefont{Scarani}},
  \bibinfo{author}{\bibfnamefont{H.}~\bibnamefont{Bechmann-Pasquinucci}},
  \bibinfo{author}{\bibfnamefont{N.~J.} \bibnamefont{Cerf}},
  \bibinfo{author}{\bibfnamefont{M.}~\bibnamefont{Du{\v{s}}ek}},
  \bibinfo{author}{\bibfnamefont{N.}~\bibnamefont{L{\"u}tkenhaus}},
  \bibnamefont{and} \bibinfo{author}{\bibfnamefont{M.}~\bibnamefont{Peev}},
  \bibinfo{journal}{Reviews of modern physics} \textbf{\bibinfo{volume}{81}},
  \bibinfo{pages}{1301} (\bibinfo{year}{2009}).

\bibitem[{\citenamefont{Xu et~al.}(2020)\citenamefont{Xu, Ma, Zhang, Lo, and
  Pan}}]{xu2020secure}
\bibinfo{author}{\bibfnamefont{F.}~\bibnamefont{Xu}},
  \bibinfo{author}{\bibfnamefont{X.}~\bibnamefont{Ma}},
  \bibinfo{author}{\bibfnamefont{Q.}~\bibnamefont{Zhang}},
  \bibinfo{author}{\bibfnamefont{H.-K.} \bibnamefont{Lo}}, \bibnamefont{and}
  \bibinfo{author}{\bibfnamefont{J.-W.} \bibnamefont{Pan}},
  \bibinfo{journal}{Reviews of Modern Physics} \textbf{\bibinfo{volume}{92}},
  \bibinfo{pages}{025002} (\bibinfo{year}{2020}).

\bibitem[{\citenamefont{Pirandola et~al.}(2020)\citenamefont{Pirandola,
  Andersen, Banchi, Berta, Bunandar, Colbeck, Englund, Gehring, Lupo, Ottaviani
  et~al.}}]{pirandola2019advances}
\bibinfo{author}{\bibfnamefont{S.}~\bibnamefont{Pirandola}},
  \bibinfo{author}{\bibfnamefont{U.~L.} \bibnamefont{Andersen}},
  \bibinfo{author}{\bibfnamefont{L.}~\bibnamefont{Banchi}},
  \bibinfo{author}{\bibfnamefont{M.}~\bibnamefont{Berta}},
  \bibinfo{author}{\bibfnamefont{D.}~\bibnamefont{Bunandar}},
  \bibinfo{author}{\bibfnamefont{R.}~\bibnamefont{Colbeck}},
  \bibinfo{author}{\bibfnamefont{D.}~\bibnamefont{Englund}},
  \bibinfo{author}{\bibfnamefont{T.}~\bibnamefont{Gehring}},
  \bibinfo{author}{\bibfnamefont{C.}~\bibnamefont{Lupo}},
  \bibinfo{author}{\bibfnamefont{C.}~\bibnamefont{Ottaviani}},
  \bibnamefont{et~al.}, \bibinfo{journal}{Adv. Opt. Photon.}
  \textbf{\bibinfo{volume}{12}}, \bibinfo{pages}{1012} (\bibinfo{year}{2020}).

\bibitem[{\citenamefont{Qi et~al.}(2007)\citenamefont{Qi, Fung, Lo, and
  Ma}}]{qi2007time}
\bibinfo{author}{\bibfnamefont{B.}~\bibnamefont{Qi}},
  \bibinfo{author}{\bibfnamefont{C.-H.~F.} \bibnamefont{Fung}},
  \bibinfo{author}{\bibfnamefont{H.-K.} \bibnamefont{Lo}}, \bibnamefont{and}
  \bibinfo{author}{\bibfnamefont{X.}~\bibnamefont{Ma}},
  \bibinfo{journal}{Quantum Info. Comput.} \textbf{\bibinfo{volume}{7}},
  \bibinfo{pages}{73} (\bibinfo{year}{2007}).

\bibitem[{\citenamefont{Zhao et~al.}(2008)\citenamefont{Zhao, Fung, Qi, Chen,
  and Lo}}]{zhao2008quantum}
\bibinfo{author}{\bibfnamefont{Y.}~\bibnamefont{Zhao}},
  \bibinfo{author}{\bibfnamefont{C.-H.~F.} \bibnamefont{Fung}},
  \bibinfo{author}{\bibfnamefont{B.}~\bibnamefont{Qi}},
  \bibinfo{author}{\bibfnamefont{C.}~\bibnamefont{Chen}}, \bibnamefont{and}
  \bibinfo{author}{\bibfnamefont{H.-K.} \bibnamefont{Lo}},
  \bibinfo{journal}{Phys. Rev. A} \textbf{\bibinfo{volume}{78}},
  \bibinfo{pages}{042333} (\bibinfo{year}{2008}).

\bibitem[{\citenamefont{Makarov}(2009)}]{makarov2009controlling}
\bibinfo{author}{\bibfnamefont{V.}~\bibnamefont{Makarov}},
  \bibinfo{journal}{New Journal of Physics} \textbf{\bibinfo{volume}{11}},
  \bibinfo{pages}{065003} (\bibinfo{year}{2009}).

\bibitem[{\citenamefont{Lydersen et~al.}(2010)\citenamefont{Lydersen, Wiechers,
  Wittmann, Elser, Skaar, and Makarov}}]{lydersen2010hacking}
\bibinfo{author}{\bibfnamefont{L.}~\bibnamefont{Lydersen}},
  \bibinfo{author}{\bibfnamefont{C.}~\bibnamefont{Wiechers}},
  \bibinfo{author}{\bibfnamefont{C.}~\bibnamefont{Wittmann}},
  \bibinfo{author}{\bibfnamefont{D.}~\bibnamefont{Elser}},
  \bibinfo{author}{\bibfnamefont{J.}~\bibnamefont{Skaar}}, \bibnamefont{and}
  \bibinfo{author}{\bibfnamefont{V.}~\bibnamefont{Makarov}},
  \bibinfo{journal}{Nature photonics} \textbf{\bibinfo{volume}{4}},
  \bibinfo{pages}{686} (\bibinfo{year}{2010}).

\bibitem[{\citenamefont{Wiechers et~al.}(2011)\citenamefont{Wiechers, Lydersen,
  Wittmann, Elser, Skaar, Marquardt, Makarov, and Leuchs}}]{wiechers2011after}
\bibinfo{author}{\bibfnamefont{C.}~\bibnamefont{Wiechers}},
  \bibinfo{author}{\bibfnamefont{L.}~\bibnamefont{Lydersen}},
  \bibinfo{author}{\bibfnamefont{C.}~\bibnamefont{Wittmann}},
  \bibinfo{author}{\bibfnamefont{D.}~\bibnamefont{Elser}},
  \bibinfo{author}{\bibfnamefont{J.}~\bibnamefont{Skaar}},
  \bibinfo{author}{\bibfnamefont{C.}~\bibnamefont{Marquardt}},
  \bibinfo{author}{\bibfnamefont{V.}~\bibnamefont{Makarov}}, \bibnamefont{and}
  \bibinfo{author}{\bibfnamefont{G.}~\bibnamefont{Leuchs}},
  \bibinfo{journal}{New Journal of Physics} \textbf{\bibinfo{volume}{13}},
  \bibinfo{pages}{013043} (\bibinfo{year}{2011}).

\bibitem[{\citenamefont{Lo et~al.}(2012)\citenamefont{Lo, Curty, and
  Qi}}]{lo2012measurement}
\bibinfo{author}{\bibfnamefont{H.-K.} \bibnamefont{Lo}},
  \bibinfo{author}{\bibfnamefont{M.}~\bibnamefont{Curty}}, \bibnamefont{and}
  \bibinfo{author}{\bibfnamefont{B.}~\bibnamefont{Qi}},
  \bibinfo{journal}{Physical Review Letters} \textbf{\bibinfo{volume}{108}},
  \bibinfo{pages}{130503} (\bibinfo{year}{2012}).

\bibitem[{\citenamefont{Braunstein and Pirandola}(2012)}]{braunstein2012side}
\bibinfo{author}{\bibfnamefont{S.~L.} \bibnamefont{Braunstein}}
  \bibnamefont{and}
  \bibinfo{author}{\bibfnamefont{S.}~\bibnamefont{Pirandola}},
  \bibinfo{journal}{Physical Review Letters} \textbf{\bibinfo{volume}{108}},
  \bibinfo{pages}{130502} (\bibinfo{year}{2012}).

\bibitem[{\citenamefont{Tamaki et~al.}(2012)\citenamefont{Tamaki, Lo, Fung, and
  Qi}}]{tamaki2012phase}
\bibinfo{author}{\bibfnamefont{K.}~\bibnamefont{Tamaki}},
  \bibinfo{author}{\bibfnamefont{H.-K.} \bibnamefont{Lo}},
  \bibinfo{author}{\bibfnamefont{C.-H.~F.} \bibnamefont{Fung}},
  \bibnamefont{and} \bibinfo{author}{\bibfnamefont{B.}~\bibnamefont{Qi}},
  \bibinfo{journal}{Physical Review A} \textbf{\bibinfo{volume}{85}},
  \bibinfo{pages}{042307} (\bibinfo{year}{2012}).

\bibitem[{\citenamefont{Wang}(2013)}]{wang2013three}
\bibinfo{author}{\bibfnamefont{X.-B.} \bibnamefont{Wang}},
  \bibinfo{journal}{Physical Review A} \textbf{\bibinfo{volume}{87}},
  \bibinfo{pages}{012320} (\bibinfo{year}{2013}).

\bibitem[{\citenamefont{Curty et~al.}(2014)\citenamefont{Curty, Xu, Cui, Lim,
  Tamaki, and Lo}}]{curty2014finite}
\bibinfo{author}{\bibfnamefont{M.}~\bibnamefont{Curty}},
  \bibinfo{author}{\bibfnamefont{F.}~\bibnamefont{Xu}},
  \bibinfo{author}{\bibfnamefont{W.}~\bibnamefont{Cui}},
  \bibinfo{author}{\bibfnamefont{C.~C.~W.} \bibnamefont{Lim}},
  \bibinfo{author}{\bibfnamefont{K.}~\bibnamefont{Tamaki}}, \bibnamefont{and}
  \bibinfo{author}{\bibfnamefont{H.-K.} \bibnamefont{Lo}},
  \bibinfo{journal}{Nature communications} \textbf{\bibinfo{volume}{5}},
  \bibinfo{pages}{3732} (\bibinfo{year}{2014}).

\bibitem[{\citenamefont{Xu et~al.}(2014)\citenamefont{Xu, Xu, and
  Lo}}]{xu2014protocol}
\bibinfo{author}{\bibfnamefont{F.}~\bibnamefont{Xu}},
  \bibinfo{author}{\bibfnamefont{H.}~\bibnamefont{Xu}}, \bibnamefont{and}
  \bibinfo{author}{\bibfnamefont{H.-K.} \bibnamefont{Lo}},
  \bibinfo{journal}{Physical Review A} \textbf{\bibinfo{volume}{89}},
  \bibinfo{pages}{052333} (\bibinfo{year}{2014}).

\bibitem[{\citenamefont{Zhou et~al.}(2016)\citenamefont{Zhou, Yu, and
  Wang}}]{zhou2016making}
\bibinfo{author}{\bibfnamefont{Y.-H.} \bibnamefont{Zhou}},
  \bibinfo{author}{\bibfnamefont{Z.-W.} \bibnamefont{Yu}}, \bibnamefont{and}
  \bibinfo{author}{\bibfnamefont{X.-B.} \bibnamefont{Wang}},
  \bibinfo{journal}{Physical Review A} \textbf{\bibinfo{volume}{93}},
  \bibinfo{pages}{042324} (\bibinfo{year}{2016}).

\bibitem[{\citenamefont{Yin et~al.}(2016)\citenamefont{Yin, Chen, Yu, Liu, You,
  Zhou, Chen, Mao, Huang, Zhang et~al.}}]{yin2016measurement}
\bibinfo{author}{\bibfnamefont{H.-L.} \bibnamefont{Yin}},
  \bibinfo{author}{\bibfnamefont{T.-Y.} \bibnamefont{Chen}},
  \bibinfo{author}{\bibfnamefont{Z.-W.} \bibnamefont{Yu}},
  \bibinfo{author}{\bibfnamefont{H.}~\bibnamefont{Liu}},
  \bibinfo{author}{\bibfnamefont{L.-X.} \bibnamefont{You}},
  \bibinfo{author}{\bibfnamefont{Y.-H.} \bibnamefont{Zhou}},
  \bibinfo{author}{\bibfnamefont{S.-J.} \bibnamefont{Chen}},
  \bibinfo{author}{\bibfnamefont{Y.}~\bibnamefont{Mao}},
  \bibinfo{author}{\bibfnamefont{M.-Q.} \bibnamefont{Huang}},
  \bibinfo{author}{\bibfnamefont{W.-J.} \bibnamefont{Zhang}},
  \bibnamefont{et~al.}, \bibinfo{journal}{Physical Review Letters}
  \textbf{\bibinfo{volume}{117}}, \bibinfo{pages}{190501}
  (\bibinfo{year}{2016}).

\bibitem[{\citenamefont{Wang et~al.}(2019)\citenamefont{Wang, Hu, and
  Yu}}]{wang2019practical}
\bibinfo{author}{\bibfnamefont{X.-B.} \bibnamefont{Wang}},
  \bibinfo{author}{\bibfnamefont{X.-L.} \bibnamefont{Hu}}, \bibnamefont{and}
  \bibinfo{author}{\bibfnamefont{Z.-W.} \bibnamefont{Yu}},
  \bibinfo{journal}{Physical Review Applied} \textbf{\bibinfo{volume}{12}},
  \bibinfo{pages}{054034} (\bibinfo{year}{2019}).

\bibitem[{\citenamefont{Lucamarini et~al.}(2018)\citenamefont{Lucamarini, Yuan,
  Dynes, and Shields}}]{lucamarini2018overcoming}
\bibinfo{author}{\bibfnamefont{M.}~\bibnamefont{Lucamarini}},
  \bibinfo{author}{\bibfnamefont{Z.}~\bibnamefont{Yuan}},
  \bibinfo{author}{\bibfnamefont{J.}~\bibnamefont{Dynes}}, \bibnamefont{and}
  \bibinfo{author}{\bibfnamefont{A.}~\bibnamefont{Shields}},
  \bibinfo{journal}{Nature} \textbf{\bibinfo{volume}{557}},
  \bibinfo{pages}{400} (\bibinfo{year}{2018}).

\bibitem[{\citenamefont{Wang et~al.}(2018)\citenamefont{Wang, Yu, and
  Hu}}]{wang2018twin}
\bibinfo{author}{\bibfnamefont{X.-B.} \bibnamefont{Wang}},
  \bibinfo{author}{\bibfnamefont{Z.-W.} \bibnamefont{Yu}}, \bibnamefont{and}
  \bibinfo{author}{\bibfnamefont{X.-L.} \bibnamefont{Hu}},
  \bibinfo{journal}{Physical Review A} \textbf{\bibinfo{volume}{98}},
  \bibinfo{pages}{062323} (\bibinfo{year}{2018}).

\bibitem[{\citenamefont{Mayers and Yao}(1998)}]{mayers1998proceedings}
\bibinfo{author}{\bibfnamefont{D.}~\bibnamefont{Mayers}} \bibnamefont{and}
  \bibinfo{author}{\bibfnamefont{A.}~\bibnamefont{Yao}}, in
  \emph{\bibinfo{booktitle}{Proceedings of the 39th Annual Symposium on
  Foundations of Computer Science (FOCS98)}} (\bibinfo{publisher}{IEEE Computer
  Society}, \bibinfo{year}{1998}), p. \bibinfo{pages}{503}.

\bibitem[{\citenamefont{Ac{\'\i}n et~al.}(2007)\citenamefont{Ac{\'\i}n,
  Brunner, Gisin, Massar, Pironio, and Scarani}}]{acin2007device}
\bibinfo{author}{\bibfnamefont{A.}~\bibnamefont{Ac{\'\i}n}},
  \bibinfo{author}{\bibfnamefont{N.}~\bibnamefont{Brunner}},
  \bibinfo{author}{\bibfnamefont{N.}~\bibnamefont{Gisin}},
  \bibinfo{author}{\bibfnamefont{S.}~\bibnamefont{Massar}},
  \bibinfo{author}{\bibfnamefont{S.}~\bibnamefont{Pironio}}, \bibnamefont{and}
  \bibinfo{author}{\bibfnamefont{V.}~\bibnamefont{Scarani}},
  \bibinfo{journal}{Physical Review Letters} \textbf{\bibinfo{volume}{98}},
  \bibinfo{pages}{230501} (\bibinfo{year}{2007}).

\bibitem[{\citenamefont{Scarani and Renner}(2008)}]{scarani2008quantum}
\bibinfo{author}{\bibfnamefont{V.}~\bibnamefont{Scarani}} \bibnamefont{and}
  \bibinfo{author}{\bibfnamefont{R.}~\bibnamefont{Renner}},
  \bibinfo{journal}{Physical review letters} \textbf{\bibinfo{volume}{100}},
  \bibinfo{pages}{200501} (\bibinfo{year}{2008}).

\bibitem[{\citenamefont{Liu et~al.}(2019)\citenamefont{Liu, Yu, Zhang, Guan,
  Chen, Zhang, Hu, Li, Jiang, Lin et~al.}}]{liu2019experimental}
\bibinfo{author}{\bibfnamefont{Y.}~\bibnamefont{Liu}},
  \bibinfo{author}{\bibfnamefont{Z.-W.} \bibnamefont{Yu}},
  \bibinfo{author}{\bibfnamefont{W.}~\bibnamefont{Zhang}},
  \bibinfo{author}{\bibfnamefont{J.-Y.} \bibnamefont{Guan}},
  \bibinfo{author}{\bibfnamefont{J.-P.} \bibnamefont{Chen}},
  \bibinfo{author}{\bibfnamefont{C.}~\bibnamefont{Zhang}},
  \bibinfo{author}{\bibfnamefont{X.-L.} \bibnamefont{Hu}},
  \bibinfo{author}{\bibfnamefont{H.}~\bibnamefont{Li}},
  \bibinfo{author}{\bibfnamefont{C.}~\bibnamefont{Jiang}},
  \bibinfo{author}{\bibfnamefont{J.}~\bibnamefont{Lin}}, \bibnamefont{et~al.},
  \bibinfo{journal}{Phys. Rev. Lett.} \textbf{\bibinfo{volume}{123}},
  \bibinfo{pages}{100505} (\bibinfo{year}{2019}).

\bibitem[{\citenamefont{Drever et~al.}(1983)\citenamefont{Drever, Hall,
  Kowalski, Hough, Ford, Munley, and Ward}}]{drever1983laser}
\bibinfo{author}{\bibfnamefont{R.}~\bibnamefont{Drever}},
  \bibinfo{author}{\bibfnamefont{J.~L.} \bibnamefont{Hall}},
  \bibinfo{author}{\bibfnamefont{F.}~\bibnamefont{Kowalski}},
  \bibinfo{author}{\bibfnamefont{J.}~\bibnamefont{Hough}},
  \bibinfo{author}{\bibfnamefont{G.}~\bibnamefont{Ford}},
  \bibinfo{author}{\bibfnamefont{A.}~\bibnamefont{Munley}}, \bibnamefont{and}
  \bibinfo{author}{\bibfnamefont{H.}~\bibnamefont{Ward}},
  \bibinfo{journal}{Appl. Phys. B} \textbf{\bibinfo{volume}{31}},
  \bibinfo{pages}{97} (\bibinfo{year}{1983}).

\bibitem[{\citenamefont{Pound}(1946)}]{pound1946electronic}
\bibinfo{author}{\bibfnamefont{R.~V.} \bibnamefont{Pound}},
  \bibinfo{journal}{Review of Scientific Instruments}
  \textbf{\bibinfo{volume}{17}}, \bibinfo{pages}{490} (\bibinfo{year}{1946}).

\end{thebibliography}

\end{document}